\documentclass[twocolumn,aps,showpacs,showkeys]{revtex4}

\usepackage{graphicx}

\newcommand{\exsitu}{{\it ex-situ} }
\newcommand{\Jone}{$J_1$\ }
\newcommand{\Jtwo}{$J_2$\ }
\newcommand{\Joneabs}{$\mid J_{1}\mid$\ }
\newcommand{\Jtwoabs}{$\mid J_{2}\mid$\ }
\newcommand{\Jc}{$J_{direct}$\ }

\begin{document}

\title{Control of interlayer exchange coupling in Fe/Cr/Fe trilayers by ion beam irradiation}

\author{S.\ O.\ Demokritov, C.\ Bayer, S.\ Poppe, M.\ Rickart, J.\ Fassbender, B.\ Hillebrands}
\address{Fachbereich Physik and Forschungs- und Entwicklungsschwerpunkt
Materialwissenschaften, Universit{\"a}t Kaiserslautern, D-67663
Kaiserslautern, Germany}

\author{D.\ I.\ Kholin and N.\ M.\ Kreines}
\address{Institute for Physical Problems Russian Academy of Science, Moscow, Russia}

\author{O.\ M.\ Liedke}
\address{Institute of Experimental Physics, University of Bialystok, Poland}

\begin{abstract}
The manipulation of the antiferromagnetic interlayer coupling in
the epitaxial Fe/Cr/Fe(001) trilayer system by moderate 5 keV He
ion beam irradiation has been investigated experimentally. It is
shown that even for irradiation with very low fluences
($10^{14}$\,ions/cm$^2$) a drastic change in strength of the
coupling appears. For thin Cr-spacers (below 0.6 -- 0.7\,nm) the
coupling strength decreases with fluence, becoming ferromagnetic
for fluences above ($2\times 10^{14}$\,ions/cm$^2$). The effect
is connected with the creation of magnetic bridges in the layered
system due to atomic exchange events caused by the bombardment.
For thicker Cr spacers (0.8 -- 1.2 \,nm) an enhancement of the
antiferromagnetic coupling strength is found. A possible
explanation of the enhancement effect is given.

\vspace*{2cm}
\end{abstract}

\pacs{}

\keywords{ion beam irradiation, interlayer exchange coupling}


\maketitle
Since the discovery of the antiferromagnetic interlayer exchange
coupling effect in the Fe/Cr/Fe layered system by Gr\"unberg et
al.\ \cite{Grunberg86} this effect has been widely investigated
both theoretically and experimentally (for a recent review see
\cite{handbook01}). Antiferromagnetically coupled layers are now
used in applications like antiferromagnetically coupled media
(AFC-media \cite{Fullerton00}) and artificial antiferromagnets,
(AAF \cite{Schmalhorst00}). In many cases such applications
greatly gain from a potential of lateral modification of the
media parameters with high resolution, after the preparation
process of the layered system has been completed. It is not
trivial to change the interlayer coupling strength after sample
preparation. Until now, to our knowledge, the only reported
methods are annealing \cite{Leng93} and charging of the spacer
with hydrogen or deuterium \cite{Klose97,Sweden,Leiner02}.
However, such techniques can hardly provide any reasonable
lateral resolution. On the contrary, beams of light ions with keV
energies known for their ability to deeply penetrate into a solid
can be focused down to 20 nm \cite{Aign98,Warin01,Lohau01}, and
should provide a promising pass to accomplish the goal. One of the
key advantages of ion irradiation is that magnetic nanopatterning
becomes feasible without a change of the sample topography. This
is especially important to avoid tribology problems in so-called
patterned media \cite{New94}.

In this Letter, we present first experimental results
demonstrating that the strength of the interlayer exchange
coupling between two ferromagnets, separated by a
non-ferromagnetic spacer, can be modified in a controlled manner
by ion beam irradiation.  It is also shown, that for some values
of the spacer thickness the ion beam bombardment {\it enhances}
the coupling.

Interaction between two magnetic layers separated by a nonmagnetic
spacer layer can be phenomenologically described by:
\begin{eqnarray}\label{eq1}
E=-J_{1}cos\phi-J_{2}cos^{2}\phi \nonumber
\end{eqnarray}
where $E$ is the magnetic coupling interface energy, $\phi$ is the
angle between the magnetizations of two magnetic layers and the
parameters \Jone and \Jtwo represent the strength of the bilinear
and biquadratic coupling, respectively \cite{Ruhrig91}. If \Jtwo
dominates and is negative, it promotes perpendicular
($90^{\circ}$) orientation of the two magnetization vectors. The
microscopic origin of the bilinear coupling is a long-range
interaction between the magnetic moments via conduction electrons
of the spacer. For smooth interfaces \Jone oscillates as a
function of the spacer thickness \cite{Parkin90, Parkin91}.
Essential roughness diminishes the bilinear coupling strength and
the amplitude of the oscillations \cite{Unguris97,Wang90}. For
perfect layered systems \Jtwo is thought to be small
\cite{Erikson93}. The experimentally observed strong biquadratic
coupling is believed to be due to extrinsic effects
\cite{Slonczewski91,Demokritov94}.

Light ion irradiation is known to be an excellent tool to modify
magnetic parameters of multilayer systems. Chappert et al.
\cite{Chappert98} have shown that ion irradiation of Co/Pt
multilayers leads to a reduction of the perpendicular interface
anisotropy. This has been attributed to an interfacial mixing of
both atom species. In FePt alloy systems an increase of the
perpendicular magnetic anisotropy due to a short range chemical
ordering has been observed after ion irradiation
\cite{Ravelosona00}. The technique has been recently applied to
exchange-bias systems, consisting of adjacent ferromagnetic and
antiferromagnetic layers. It was shown that the magnitude and
direction of the exchange-bias field can be tailored by ion
irradiation if a magnetic field is applied during bombardment
\cite{Mewes00,Mougin01}.


Epitaxial Fe/Cr/Fe(001) samples used in the current studies  were
prepared in an ultra high vacuum molecular-beam epitaxy system
with the base pressure below $5\times10^{-11}$\,mbar. A Cr buffer
with a thickness of 100\,nm providing a lattice matched template
for the subsequent growth of the Fe/Cr/Fe \cite{Fullerton93}
system was deposited on a MgO(001) substrate.  Two Fe films
separated by an wedge-shape, 0.4 to 4\,nm thick Cr spacer were
deposited on the buffer. Different samples with the thickness of
the Fe films from 5 to 10\,nm have been prepared. Finally the
system was covered by 3\,nm Cr to avoid corrosion for \exsitu
measurements. The details of substrate preparation and the growth
procedure are published elsewhere \cite{Rickart01}. Figure 1
displays the topography of the lower Fe film and of the Cr spacer
as observed by STM. Atomic terraces and monoatomic steps are
clearly seen in the images.

\begin{figure}[htbp]
\begin{center}
\scalebox{1}{\includegraphics[width=8.5 cm,clip]{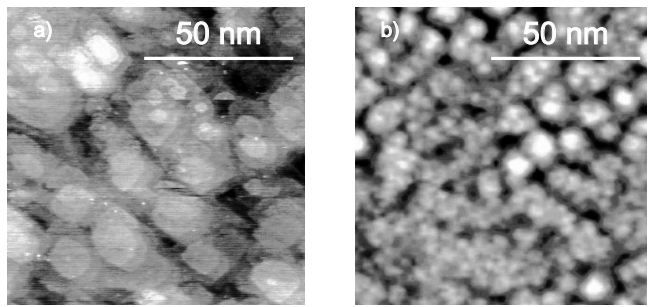}}
\end{center}
\caption{STM images of the film surfaces: a) bottom Fe film,
z-scale 1.3\,nm, RMS = 0.11 nm; b) Cr spacer, z-scale 1.5\,nm, RMS
= 0.18\,nm.}
\end{figure}

Using the measured RMS values of the surface roughness (0.11 and
0.18 nm for the Fe and the Cr surface, respectively) and assuming
uncorrelated thickness fluctuations for the two films, one
obtains an RMS value for the thickness fluctuations of the Cr
spacer of 0.14 nm, which is close to the thickness of one
monolayer (ML). Based on this value and assuming a Gaussian
distribution of the probability for the spacer to consist of a
given number of monolayers, one obtains, for example, for a
nominal thickness of the Cr spacer, $d_{Cr}$ of 4 ML that: 38\% of
the film area has $d_{Cr}$ = 4 ML, 24\% has $d_{Cr}$ = 3 ML, 6\% 2
ML, 0.6\% 1 ML, and 0.025\% corresponds to direct contact between
the two Fe films (so-called "magnetic bridges"). The latter
provide a strong ferromagnetic coupling between the Fe films.

Irradiation was performed with 5\,keV He$^{+}$ ions without
applied magnetic field with the sample being kept at room
temperature. TRIM simulations \cite{Ziegler85} show that for the
used parameter set most ions pass both magnetic layers and are
stoped in the Cr buffer layer.  The maximum fluence used was
$8\times 10^{14}$\,ions/cm$^2$. The interlayer coupling was
derived from the magnetization curves recorded by longitudinal
magneto-optical Kerr-effect (MOKE) magnetometry. A magnetic field
of up to 6 kOe was applied in the plane of the sample parallel
either to the easy or to the hard magnetic axes of the four-fold
magnetic anisotropy of the Fe(001) films.

The magnetization curves measured for the field applied along the
easy [100]-axis show several jumps, characteristic for magnetic
double layers with antiferromagnetic and $90^{\circ}$-coupling
\cite{Ruhrig91}. The saturation field, $H_{S}$, which is
proportional to  $\mid J_1+2J_2\mid$ extracted from the
magnetization curves is shown in Fig.\ 2 as a function of the
nominal Cr-spacer thickness for different ion irradiation
fluences.  The data obtained on the as-prepared sample clearly
demonstrate both long- and short-period oscillations with a
moderate amplitude in agreement with the RMS value of the spacer
thickness fluctuation obtained from the STM studies. The arrows
indicate the first three oscillation maxima of the coupling
strength. As it is seen in Fig.\ 2, such a well prepared layered
magnetic system is very sensitive to ion irradiation. The first
oscillation maximum ($d_{Cr}=$ 0.58 nm = 4 ML) exhibits the
strongest effect of the irradiation on the coupling strength.
Even the lowest used ion fluence of $0.5\times
10^{14}$\,ions/cm$^2$ reduced $H_{S}$ nearly by 25\%. For the
fluences above $2\times 10^{14}$\,ions/cm$^2$ no antiferromagnetic
coupling is observed for this thickness of the Cr-spacer. The
change of the measured coupling strength for thicker Cr-spacers
is more intriguing: the coupling {\it increases} for small ion
fluences and then decreases for fluences above $1\times
10^{14}$\,ions/cm$^2$ \cite{note}.

\begin{figure}[htbp]
\begin{center}
\scalebox{1}{\includegraphics[width=9 cm,clip]{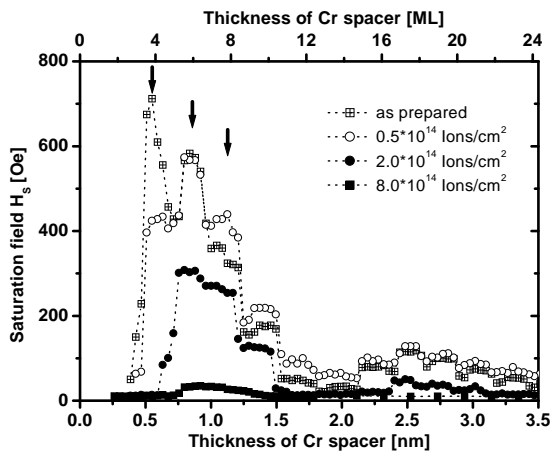}}
\end{center}
\caption{Saturation field $H_S$ as a function of the Cr spacer
thickness for the 10 nm Fe/$d_{Cr}$ Cr/10 nm Fe trilayer system
as prepared and after the irradiation with the fluences as
indicated. The arrows indicate the first three maxima of the
interlayer coupling.}
\end{figure}

An additional study made by means of Brillouin light scattering on
spin waves has indicated no change in the four-fold in-plane and
out-of-plane anisotropy constants after the bombardment for the
studied fluence range.

Of particular interest are the dependencies of the coupling
constants \Jone and \Jtwo as functions of ion fluence and spacer
thickness. From the measured remagnetization curves the fluence
dependence of those constants for the spacer thicknesses
corresponding to the first (4 ML), second (6 ML), and third (8
ML) oscillation maxima have been evaluated.  Note that only the
values of the antiferromagnetic ($ J_{1}<0$) and
$90^{\circ}$-degree ($J_{2}<0$) coupling constants can be usually
derived in such a way.  The data is presented in Fig.\ 3.  It is
clearly seen from the figure, that \Joneabs strongly decreases
with the fluence for $d_{Cr}$= 4 ML, while it shows  a maximum
for fluences near $0.5\times 10^{14}$\,ions/cm$^2$ for $d_{Cr}$
equal to 6 and 8 ML.  \Jtwoabs instead shows a monotonic
decrease.  Thus, one can conclude from Fig.\ 3 that the increase
of the saturation field at small irradiation fluences is caused
by the increase of \Joneabs.

\begin{figure}[htbp]
\begin{center}
\scalebox{1}{\includegraphics[width=9 cm,clip]{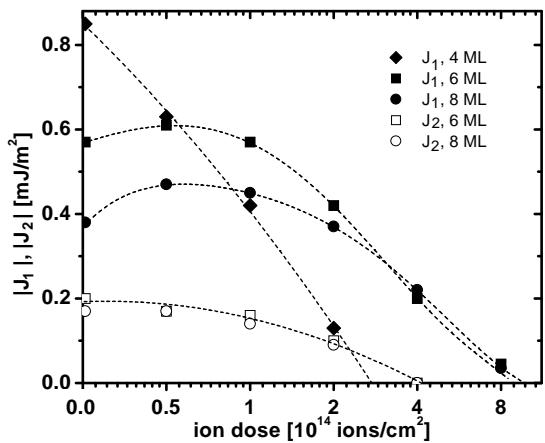}}
\end{center}
\caption{$J_1$ and $J_2$ obtained form the remagnetization curves
for different maxima versus the fluence. The lines are guides to
the eye.}
\end{figure}


The origin of the observed phenomena is not understood in all
details yet, but they are definitely connected with the surface
intermixing caused by the He ions. To understand this qualitively
let us first consider the fluence dependence of \Joneabs for the
nominal thickness $d_{Cr}$= 4 ML. Direct magnetic bridges between
the two Fe films provide strong direct ferromagnetic coupling, \Jc
$\approx 2A/d_{ML}$= 280 mJ/m$^2$, where $A = 2\times 10^{-11}$
J/m is the bulk exchange constant of Fe and $d_{ML}$ = 0.144 nm
is the thickness of one monolayer. The contribution to the
measured interlayer coupling due to the magnetic bridges can be
then easily estimated, since the percentage area of those bridges
is known from the RMS analysis of the STM images discussed above.
For the as-prepared sample the obtained value of 0.07 mJ/m$^2$ is
essentially smaller than the measured one and indicates that the
interlayer coupling via the spacer is an order of magnitude
larger than direct coupling across the bridges.

It is known that an ion propagating within a lattice partly
dissipates its energy due to nuclear collisions \cite{Ziegler85}.
Such collisions cause recoil of atoms of the lattice and creation
of lattice defects and intermixing. Although light ions, like
helium, have a low displacement rate and cause a short range
intermixing, these processes are of importance if taking place at
the interface. Estimations show that a 5 keV He ion initiates in
average between one and two atom pair exchange events per
monolayer in Fe and Cr \cite{Urbassek}. In the areas with
$d_{Cr}$ = 1 ML such an exchange event induces an atomic size
magnetic bridge. Assuming that each ion generates one exchange
event per monolayer as a lower bond and taking into account also a
possibility of two successive exchange events at adjacent lattice
sites, one obtains that for the fluence of $2\times
10^{14}$\,ions/cm$^2$ the relative area of the magnetic bridges
increases to 0.2\% and their contribution to the interlayer
coupling is 0.6 mJ/m$^2$. This is in rather good agreement with
the experimentally observed coupling reduction of 0.72 mJ/m$^2$.
The calculation also demonstrates that the probability for
formation of magnetic bridges due to the bombardment decreases
exponentially with the nominal spacer thickness at a given
interface roughness. Thus, it is not surprising that the effect
of the irradiation is weaker for larger spacer thicknesses (the
second and third maximum).

Surprising is, however, the observation of an {\it increase} of
the antiferromagnetic coupling strength. These findings might be
related to the fact, that, first, an intermixing at the Fe/Cr(001)
interface with a width of 1-2 monolayers is supposed to be
energetically favorable \cite{Sauer96,Freyss97,Heinrich99}, but
it is usually not completely achieved during the film growth
because of kinetic growth effects. Second, He ions in the
discussed energy range very effectively transfer energy to
phonons (8 -- 12 eV per monolayer), which in turn help the system
to relax into this optimum.

It is known that the interlayer coupling in the Fe/Cr/Fe(001)
layered system can be increased by a gentle annealing
\cite{Leng93}. Stronger coupling in this case is usually
connected with higher lateral homogeneity of interface
intermixing between Fe and Cr. On the other hand, a close
relation between a homogeneous intermixing at the interface and
the interlayer coupling has been recently nicely demonstrated for
Fe/Si/Fe \cite{Gareev02}: the introduction of two monolayers of
Fe$_{0.5}$Si$_{0.5}$ at every Fe/Si interface brought about a
much stronger coupling, as observed on the samples where the
intermixing took place naturally.

Using all the above presented facts, the observed increase of the
interlayer coupling can be qualitatively understood as the effect
of "phonon annealing": An ion propagating in the lattice creates
pulses of hyperthermic phonons along its trajectory. The emitted
phonons increase the probability that those parts of the
interfaces, where the energetically favorable mixing has not been
reached during the growth, move towards this equilibrium. Note
here, that since this process at its end can produce additional
energy, the phonons do not spend their energy and act just as a
catalyst. In a similar way as it is observed in the Fe/Si/Fe
system \cite{Gareev02}, a higher degree of the interface
homogeneity causes higher interlayer coupling. The proposed model
is rather speculative and demands further studies, which are
outside the scope of this paper.

In conclusion, we have experimentally shown that
antiferromagnetic interlayer coupling of the Fe/Cr/Fe(001)
layered system can be modified using ion beams after system
preparation. Depending on the thickness of the Cr-spacer and the
ion beam fluence the coupling strength can either decrease or
increase. Our results might open new fields of applications of
antiferromagnetically coupled systems by laterally tayloring the
coupling strength with the potential high lateral resolution of
ion beams. Systems with controlled spatial variation of the local
magnetization can be fabricated using this approach.

Support by the Deutsche Forschungsgemeinschaft is gratefully
acknowledged. C.B. acknowledges support by the Studienstiftung
des Deutschen Volkes. The authors are also indebted to H.
Urbassek for his fruitful comments on the ion intermixing power.



\begin{thebibliography}{99}
\bibitem{Grunberg86} P.\ Gr\"unberg, R.\ Schreiber, Y.\ Pang, M.\ B.\ Brodsky and H.\ Sowers,
                        Phys.\ Rev.\ Lett.\ {\bf 57}, 2442 (1986).
\bibitem{handbook01} D.E.\ B\"urgler, S.\ O.\ Demokritov, P.\ Gr\"unberg, M.T.\
                        Johnson, Handbook of Magnetic Materials, vol 13, Ed.\ K.J.H.\ Buschow, Elsevier,
                        Amsterdam, 2001.
\bibitem{Fullerton00} Eric E.\ Fullerton, D.\ T.\ Margulies, M.\ E.\ Schabes,
                        M.\ Carey, B.\ Gurney, A.\ Moser, M.\ Best, G.\ Zelter, K.\ Rubin, and H.\
                        Rosen, M.\ Doerner, Appl.\ Phys.\ Lett.\ {\bf 77}, 3806
                        (2000).
\bibitem{Schmalhorst00} J.\ Schmalhorst, H.\ Br\"uckl, and G.\ Reiss, R.\
                        Kinder, G.\ Gieres, and J.\ Wecker, Appl.\ Phys.\ Lett.\ {\bf 77}, 3456
                        (2000).
\bibitem{Leng93}        Q.\ Leng, V.\ Cross, R. Sch\"afer, A.\
                        Fuss, P.\ Gr\"unberg, and W.\ Zinn, Journ.\ Mag.\ Mag.\ Mat. {\bf
                        126},  367 (1993).
\bibitem{Klose97} F.\ Klose, Ch.\ Rehm, D.\ Nagengast, H.\ Maletta and A.\ Weidinger,
                        Phys.\ Rev.\ Lett.\ {\bf 78}, 1150 (1997).
\bibitem{Sweden} B.\ Hj\"ovarsson, J.\ A.\ Dura, P.\ Isberg, T.\ Watanabe, T.\
                    J.\ Udovic, G.\ Andersson, and C.\ F.\ Majkrzak, Phys.\ Rev.\ Lett.\ {\bf 79},
                    901 (1997).
\bibitem{Leiner02} V.\ Leiner, M.\ Au, T.\ Schmitte, H.\ Zabel,
                    Appl.\ Phys.\ A, in print 2002.
\bibitem{Aign98} T. Aign, P. Meyer, S. Lemerle, J. P. Jamet, J.
                    Ferre, V. Mathet, C. Chappert, J. Gierak, C. Vieu, F. Rousseaux,
                    H. Launois, H. Bernas, Phys. Rev. Lett. {\bf 81}, 5656 (1998).
\bibitem{Warin01} P. Warin, R. Hyndman, J.N. Chapman, J. Ferre,J. P. Jamet, V. Mathet , C. Chappert,
                    Journ. Appl. Phys. {\bf
                    90}, 3850 (2001).
\bibitem{Lohau01} J. Lohau, A. Moser, C. T. Rettner, M. E. Best,
                    B. D. Terris, Appl. Phys. Lett. {\bf 78}, 990 (2001).
\bibitem{New94} R. M. H. New, R. F. W. Pease, R. L. White, J. Vac.
                    Sci. Technol. B  {\bf 12}, 3196 (1994).
\bibitem{Ruhrig91} M.\ R\"uhrig, R.\ Sch\"afer, A.\ Hubert, R.\ Mosler, J.\
                        A.\ Wolf, S.\ Demokritov, and P.\ Gr\"unberg, Phys.\ Stat.\ Sol.\ (a) {\bf 125}, 635
                    (1991).
\bibitem{Parkin90} S.\ S.\ P.\ Parkin, N.\ More, and K.\ P.\ Roche, Phys.\ Rev.\
                    Lett.\ {\bf 64}, 2304 (1990).
\bibitem{Parkin91} S.\ S.\ P.\ Parkin, Phys.\ Rev.\ Lett.\ {\bf 67}, 3598 (1991).
\bibitem{Unguris97} J.\ Unguris, R.\ J.\ Celotta, and D.\ T.\ Pierce, Phys.\
                        Rev.\ Lett.\, {\bf 79} 2734 (1997).
\bibitem{Wang90} Y.\ Wang and P.\ M.\ Levy, J.\ L.\ Fry, Phys.\ Rev.\ Lett.\
                            {\bf 65}, 2732 (1990).
\bibitem{Erikson93} R.\ P.\ Erikson, Kristl B.\ Hathaway and James R.\ Cullen,
                    Phys.\ Rev.\ B {\bf 47}, 2626 (1993).
\bibitem{Slonczewski91} J.\ C.\ Slonczewski, Phys.\ Rev.\ Lett.\ {\bf 67},3172
                    (1991); J.\ C.\ Slonczewski, Journ.\ Mag.\ Mag.\ Mat.\ {\bf 150}, 13
                    (1995).
\bibitem{Demokritov94} S.\ Demokritov, E.\ Tsymbal, P.\ Gr\"unberg, W.\ Zinn, I.\ K.\ Schuller, Phys.\ Rev.\ B {\bf 49},
                         720 (1994).
\bibitem{Chappert98} C.\ Chappert, H.\ Bernas, J.\ Ferre, V.\ Kottler, J.-P.\
                        Jamet, Y.\ Chen, E.\ Cambril, T.\ Devolder, F. Rousseaux, V.\ Mathet, and H.\
                        Launois, Science {\bf 280}, 1919 (1998).
\bibitem{Ravelosona00} D.\ Ravelosona, C.\ Chappert, and V.\ Mathet, H.
                        Bernas, Appl.\ Phys.\ Lett.\ {\bf 76}, 236
                        (2000).
\bibitem{Mewes00} T.\ Mewes, R.\ Lopusnik, J.\ Fassbender, B.\ Hillebrands, M.\ Jung,
                        D.\ Engel, A.\ Ehresmann, H.\ Schmoranzer Appl.\ Phys.\ Lett.\ {\bf76}, 1057
                        (2000).

\bibitem{Mougin01} A.\ Mougin, T.\ Mewes, M.\ Jung, D.\ Engel,
                    A.\ Ehresmann, H.\ Schmoranzer, J.\ Fassbender, B.\ Hillebrands,
                    Phys. Rev. B\ {\bf 63}, 060409(R) (2001).
\bibitem{Rickart01} M.\ Rickart, B.\ F.\ P.\ Roos, T.\ Mewes, J.\ Jorzick, S.\ O.\ Demokritov, B.\ Hillebrands,
                        Surf.\ Sci.\ {\bf 495}, 68 (2001).
\bibitem{Fullerton93} E.\ E.\ Fullerton, M.\ J.\ Conover, J.\ E.\ Mattson, C.\ H.\ Sowers,
                       and S.\ D.\ Bader, Appl.\ Phys.\ Lett.\ {\bf 63}, 1699 (1993).

\bibitem{Ziegler85} J.\ Ziegler, J.\ Biersack, and U.\ Littmark, The Stopping
                        of Ions in Matter, Pergamon, New York, 1985
\bibitem{note}      Slightly varying the preparation conditions, the values of the RMS roughness as well as the
                    interlayer coupling constants and the oscillation amplitude can be changed.
                    For simplicity only the data obtained on one and the same sample are
                    presented in the paper.

\bibitem{Urbassek}    H. Urbassek, private communication.

\bibitem{Heinrich99}  B.\ Heinrich, J.\ F.\ Cochran, T.\
                        Monchesky, and R.\ Urban, Phys. Rev. B\ {\bf 59}, 14520
                        (1999).
\bibitem{Sauer96}     Ch.\ Sauer, F.\ Klinkhammer, E.\ Yu.\ Tsymbal, S.\ Handschuh, Q.\ Leng and
                        W.\ Zinn, Journ.\ Mag.\ Mag.\ Mat. {\bf 161}, 49 (1996).
\bibitem{Freyss97}      M.\ Freyss, D.\ Stoeffler, and H.\ Dreysse, Phys. Rev. B\ {\bf 56}, 6047 (1997).
\bibitem{Gareev02}     R.R.\ Gareev, D.\ E.\ B\"urgler,  M.\ Buchmeier, R.\ Schreiber, and P.\
                       Gr\"unberg, submitted to Appl.\ Phys.\ Lett.



\end{thebibliography}
\end{document}